\documentclass[12pt,a4paper]{article}
\usepackage{graphicx}
\usepackage{cite}
\usepackage{latexsym}
\usepackage{url}
\usepackage{geometry}
\usepackage{datetime}
\newdate{date}{15}{10}{2017}
\date{\displaydate{date}}

\title{AI in Software Engineering: \\Case Studies and Prospects
}
\author{
Lei Wang\thanks{The author conducted this work while enrolled as a master's student at UWA, specifically for the CITS5502 Software Processes unit in 2017.}\\
University of Western Australia}

\begin{document}

\maketitle

\begin{abstract}

Artificial intelligence (AI) and software engineering (SE) are two important areas in computer science. In recent years, researchers are trying to apply AI techniques in various stages of software development to improve the overall quality of software products. Moreover, there are also some researchers focus on the intersection between SE and AI. In fact, the relationship between SE and AI is very weak; however, methods and techniques in one area have been adopted in another area. More and more software products are capable of performing intelligent behaviour like human beings. In this paper, two cases studies which are IBM Watson and Google AlphaGo that use different AI techniques in solving real world challenging problems have been analysed, evaluated and compared. Based on the analysis of both case studies, using AI techniques such as deep learning and machine learning in software systems contributes to intelligent systems. Watson adopts `decision making support' strategy to help human make decisions; whereas AlphaGo uses `self-decision making' to choose operations that contribute to the best outcome. In addition, Watson learns from man-made resources such as paper; AlphaGo, on the other hand, learns from massive online resources such as photos. AlphaGo uses neural networks and reinforcement learning to mimic human brain, which might be very useful in medical research for diagnosis and treatment. However, there is still a long way to go if we want to reproduce human brain in machine and view computers as thinkers, because human brain and machines are intrinsically different. It would be more promising to see whether computers and software systems will become more and more intelligent to help with real world challenging problems that human beings cannot do.

\end{abstract}


\section{Introduction}

Artificial intelligence (AI) and software engineering (SE) are two important areas in computer science. Software development process is a complex process which needs intensive human activities to design and build high quality software~\cite{shankari2014}. Moreover, developers should have enough knowledge in both the problem domain and programming domain~\cite{shankari2014}. On the other hand, AI focuses on building intelligent software systems to allow machine to learn, reason, act and perceive~\cite{pawar2016}. In recent years, researchers are trying to apply AI techniques in various stages of software development to improve the overall quality of software products. Moreover, there are also some researchers focus on the intersection between SE and AI. In fact, the relationship between SE and AI is very weak; however, methods and techniques in one area have been adopted in another area. Many techniques and methodologies have been discovered from both areas such as Computational Intelligence (CI), Ambient Intelligence (AI), Knowledge-based Software Engineering (KBSE) and Agent oriented Software Engineering (AOSE)~\cite{jain2011}. 

Many AI applications have been reported in the history. In 1951, the first AI based program was written. After 4 years, the first self-learning game playing program was created successfully. In 1961, the first robot was introduced into General Motors Assembly line. Three years later, An AI program was demonstrated to have the ability to understand natural language. In 1965, the first chat robot Eliza was invented. In 1974, the first autonomous vehicle is created at Stanford AI lab. In 1989, A neural network was introduced in the first autonomous vehicle. In 1997, IBM deep beats Garry Kasparov at chess. In 1999, the first emotional AI is demonstrated in MIT AI lab. The concepts of many current popular AI applications are not new to us like self-learning game playing program, question answer robot and self-driving vehicles. However, the techniques or the deep algorithms used are evolving. 

AI techniques have been widely used in many areas. For example, in 2009, the first self-driving car successfully incorporated AI techniques like image recognition, machine learning and deep learning in its software systems. In recent years, AI assistant on mobile phones and computers like Siri and Cortana has become a mainstream to help people deal with daily tasks intelligently. Search engines and translation systems like Google also adopts machine learning techniques and cloud computing services instead of using the traditional methods. With deep learning, face, voice and image recognition also allow users to interact with mobile devices and computers in a more natural way. IBM Watson beats the two best champions in real-time competition~\cite{ferrucci2012}. DeepQA in Watson adopts many algorithms to analyse information in different dimensions like source reliability, geography and time. It trains questions and answers to learn, features and confidence values are generated after this training process~\cite{ferrucci2012}. In contrast, Google AlphaGo uses a combination of artificial neural networks and tree search in mastering the game of Go~\cite{silver2017}. AI techniques used in AlphaGo are trying to mimic human brain to make smarter decisions which contribute to the best results. AI techniques have been widely used in many areas in people's daily life, which promotes further studies on using AI in software systems.

\section{Literature Review}

Similar to other areas, the quality of software improves gradually based on the developers' knowledge, past experience and expertise~\cite{meziane2009}. Significant progress has been made in software methodologies, development environments and programming languages since then. However, challenging issues still exist since the natural complexity of software itself. It is still hard to predict the delivery of software products and the overall budget of software development. In addition, High rate of software failures still appear in SE. AI has shown visible growth which affects software development recently~\cite{meziane2009}. More and more researchers are trying to apply AI techniques to enhance and support SE so that high quality software can be built within limited budget and time. AI techniques such as Case-based Reasoning (CBR), Knowledge-based System (KBS), neural networks (NNs) and fuzzy logic have been used in many stages of software development to help improve the quality of software. 

AI has become a popular area in recent years which enhances human experience in many fields such as service systems and manufacturing~\cite{pannu2015}. AI aims at creating software systems which are capable of executing tasks like human beings~\cite{basu2017}. For example, AI systems can help write and execute codes instead of manually writing codes and test cases. Moreover, these AI systems can continuously evolve and improve based on the learning process from human inputs. In recent years, researchers have tried to explore the role of these intelligent systems in software development process~\cite{basu2017}. Massive research in AI area contributes to the growing techniques like expert systems, and these expert systems have been widely used in science, business, engineering and medicine to deal with challenging problems~\cite{pannu2015}. However, AI is not the same as psychology since it focuses on analysis and computations, and it is also different from computer science since it concentrates on reasoning, learning and gathering information~\cite{pannu2015}. Therefore, machines using AI techniques like artificial neural networks and natural language processing (NLP) are much smarter and more useful in helping with real world challenging problems~\cite{pannu2015}. 

Many applications of AI techniques in SE have been reported in the literature. There are some major AI techniques used in software development to help improve the performance and user experience of software products~\cite{kulkarni2016}. Firstly, intelligent agents allow developers to handover their tasks to smart agents. These agents can help developers discover ignored information, give suggestions and analyse complex data intelligently~\cite{russell2010}. Secondly, machine learning has been widely used in speech recognition, web searching and self-driving cars. It allows the machine to learn by itself without the need of explicitly programming~\cite{hinton2007}. Thirdly, reasoning and knowledge representation can assist developers in designing and building complex software systems much easier~\cite{sowa1992}. Fourthly, statistical model provides assumptions with probability distributions based on the given data~\cite{kulkarni2016}. Fifthly, search and optimization can be used in finding the most cost effective and high performance outcomes based on the given constraints~\cite{akbari2011}. Sixthly, fuzzy logic can be used in control systems and embedded micro controllers to give a clear decision based on unclear input information~\cite{zadeh1965}. Last but not least, Turing test is used to check whether the machine can perform intelligent behaviour like human beings~\cite{turing1950}. These AI techniques have shown great success in helping software development to improve the quality of software products and thus enhance user experience.

AI techniques have been applied in various stages in software development process. During project initialization stage, KBS can be used for project planning and effort estimation. Since we learn from past projects, we will gain the abilities and use what we have learned from previous projects to improve future projects and KBS attempts to use previous experiences in planning new software products~\cite{meziane2009}. In addition, NNs have been used in risk analysis and outcome predictions~\cite{meziane2009}. Some key features for risk analysis need to be identified based on previous projects, and then these features are passed to NNs for training. After that, for new project, NNs can predict whether the project will be failed, partially failed or succussed. AI techniques can also help with requirements engineering and software design. Since requirements are sometimes incomplete, ambiguous and difficult to manage, KBS and ontologies can be used in modelling the problem domain and managing software requirements~\cite{meziane2009}. Apart from that, using expert systems in designing software based on stored design patterns and functionalities plays an important role in software development. These AI techniques can ensure higher efficiency and quality of software products in software processes~\cite{meziane2009}. Effective software testing strategies are also very important during software development process. Many AI techniques such as Genetic Algorithms (GAs) and AI planning have been used to generate optimal test cases~\cite{sorte2015}. Since manual test for user interfaces might be inadequate, GAs can generate potential tests by modelling user actions as genes, and GAs can mutate test cases provided by testers to ensure the effectiveness of testing~\cite{meziane2009}. AI planning, on the other hand, can mutate the plans to mimic some potential defects in software systems. Moreover, AI planning can check the state transition of software operations based on the input of action sequences. 

Software development needs intensive activities conducted by skilled human with enough relevant knowledge and experiences~\cite{pawar2016}. Therefore, AI techniques such as expert systems are important in assisting most of activities in software development process. Moreover, using AI techniques, many challenging issues in SE can be relieved and high quality software products can be produced. In this research project, two cases studies which are IBM Watson and Google AlphaGo that use AI techniques in solving real world challenging problems will be analysed, evaluated and compared.

\section{AI techniques in Software Development}

Two case studies which are IBM Watson and Google AlphaGo have been analyzed in the following sections. The DeepQA used in Watson is a large-scale natural language processing (NLP) system that trains through machine learning techniques based on a large number of analytics. DeepQA has become a famous software architecture for analysis and reasoning in AI area~\cite{ferrucci2010}. In contrast, Google AlphaGo uses artificial neural networks to mimic human brain to learn from massive data and make decisions intelligently. 

\subsection{Case study 1: IBM Watson}
It was February 2011, in Jeopardy!, the most popular American quiz competition television game show, a new king was born, by defeating the most famous two winners in its history with an absolute ahead~\cite{baughman2014}. Especially the new king was no longer a human but a computer system named Watson, which attracted people's attention to the artificial intelligence Q \& A system. With the diversification of products and requirements, higher labor costs and quality of customer service are asked. Moreover, the static FAQs (frequently asked questions) method is difficult to give users a quick and accurate solution, manual service often repeats some same and simple solutions~\cite{asakiewicz2017}. With AI technology, the pressure on customer services can be reduced and even give better decision beyond human ability.

The understanding of customer questions in traditional system is based on certain rules or keywords that always leads to overly dependent on search features and documentation~\cite{liu2009}. But this is obviously only applicable to common problems, without flexibility and particularity. But with AI technology, the system can be more specific in analysis of the field and meaning contained in these questions, to achieve deep understanding and not just simply extract keywords.
The answer in the traditional system relies on the content deployed to the system in advance, and once a new problem arises, the traditional system cannot give a new effective solution~\cite{liu2009}. But with AI technology, which gives the system an ability to learn, continuous learning from different but similar cases will gradually come to an increasingly efficient solution. What's more, human are quite keen to find a system that covers knowledge base from all areas with super computing power that can exceed the human limits. This kind of system can find hidden relationships based on self-learning and put forwards more valuable warnings and solutions to new problems.

The DeepQA uses a parallel and probabilistic model, which includes using the NLP to decompose and analyse the questions. This is not only the extraction of the keywords, but also the classification of the question types based on these keywords. Then we determine several possible fields of the answer, and search answer parallelly in these fields, through the scoring to arrange the most reasonable answer~\cite{ferrucci2010}. With the expansion of the knowledge base and the deepening of training, the system will have its own best confidence value in the question. The DeepQA system uses machine learning to calculate the confidence value: the engineers prepared a set of questions with correct answer and several similar but wrong answers, and DeepQA ranked these answers itself. Then they give DeepQA the confidence value of these answers and adjust the parameters to narrowing the gap. By repeating this training process until they get the best model. The DeepQA system has formed several parallel assumptions from the beginning, and these assumptions are based on probability information from questions, searching and answering~\cite{baughman2014}. This makes the error can be constantly corrected, and the use of self-confidence can significantly reduce the costs of computation. The DeepQA of Watson is very successful, with a mass knowledge storage from all over the world, a great computing ability far beyond human ability, and even has an excellent learning ability. However, it lacks the most important human ability: thinking. This happened significantly after learning the "urban dictionary", a popular language knowledge dataset contains many uncivilized phrases, the system began to burst foul and it was so frequently and seriously that the researchers had to remove the dataset and added language Filtering function.

Optimization is a very important stage in the software development process. On the one hand, optimization meets the new requirements feedback by customers; on the other hand, the bugs we cannot predict or find in the front various stages will occur in this stage. From Watson's foul, we can further discover deeper issues that are of great influence on the software development. Watson expands its knowledge bases greedily and continue learning from these bases and evolution, but it cannot understand the content. The system can answer the intuitive question raised by human quickly and accurately, where it does better than human obviously, but it is harder or impossible now to understand the reason why people asked this question, what is the meaning of this question and even the meaning of its own answer. That is, Watson can only deal with the character symbols and cannot think about the meaning of the content, the feelings of human and the deep meaning in communication, which is clearly an extremely important requirement for customer service, and also the most important and difficult points of the optimization process of Watson.

With the database of Watson continues to expand, on the one hand, the database provides a more comprehensive reference; on the other hand, it also makes Watson face more similar and chaotic information, which leads to a lot of retrieval and judgment. As a result, Watson's work is efficiency reduced, and it even makes more mistakes in the field of cross-disciplinary. At the same time, the different forms of question expression are often supported by different evidence, sometimes leading to the relevant but maybe completely incorrect answer. Especially Watson is applied in the medical industry; such mistakes are certainly not allowed. There is still a long way to further improve the accuracy rate, and achieve optimization.

\subsection{Case study 2: Google AlphaGo}

Recently, Google's AI subsidiary DeepMind, which created the first computer program that defeated a professional human Go player has unveiled their latest evolution of grandmaster-beating Go-playing AI - AlphaGo Zero. They have published a paper in the top scientific journal nature to demonstrate the technology they used in AlphaGo Zero~\cite{silver2017}. The strategy board game Go originated in ancient China 2,500 years ago, despite its simple rules, it is a game of profound complexity, even more difficult than chess~\cite{gowiki2017}. The enormous search space (Go board has $19 \times 19$ grid of lines, compared with the chess board has only $8 \times 8$ grids) and the difficulty of evaluating the moves and positions make the implementation of artificial intelligence in Go game has often been viewed as the most challenging task. The way of developing a Go AI is much different from other game AIs with a relatively low complexity. It is possible to find a simple and clever algorithm for a small-scale AI problem like the A* algorithm for shortest path finding. However, using conventional approaches to implement Go AI seems not to be feasible. 

The easiest way of solving an artificial intelligence problem is to mimic the behaviour of human proficient counterparts. The first truly successful form of artificial intelligence system in the history - expert system works in this way~\cite{expertwiki2017}. However, Go is played primarily through feel and intuition~\cite{deepmind2017}. A professional human Go player sometimes make relatively easier judgments based on their instincts, but the instincts about the territorial advantage and a position's strength are difficult to quantify. If we cannot precisely transform human feel and intuition into detailed instructions and rules, we will not be able to implement the computer system accordingly. Due to another fact that a precise prediction is almost impossible, if a human player wants to see three steps ahead at the beginning of a Go game, we have at most $(19 \times 19) \times (19 \times 19 -1) \times (19 \times 19 -2) = 46,655,640$ possible situations to consider~\cite{chen2016}, which is infeasible in a real world game. Therefore, another often used strategy for computer programs - brutal force will not work in Go AI, because the full prediction of a Go game even without considering the ``capture'' will still have $(19 \times 19)! = 351!$ possible samples. 

DeepMind introduced the state-of-the-art artificial intelligence algorithms in the development process of their Go AI -  AlphaGo. Unlike its predecessor, AlphaGo Zero taught itself to master the Go game without learning (distracting) from any human players, just as its name it learned from zero. They use reinforcement-learning algorithm to train AlphaGo Zero, the whole training process started and continued without any human intervention for about 3 days~\cite{gowiki2017}. The face-off between the old version AlphaGo that beat world champion and its successor AlphaGo Zero ended up with the result: AlphaGo Zero won 100-0. 

Historically, AI was, and in some extent still is, a domain of reproducing aspects of human intelligence and thereby solving problems without human intervention. The ultimate goal of AI is called strong AI, for a strong AI system, only the ability to interact with the environment and the ability to learn from the interactions have to be given. All the other common sense and knowledge would gain as time passed by~\cite{strongai2017}. The most excited thing about AlphaGo Zero is that, in some extent, it is a little bit like strong AI in the world of Go game. AlphaGo Zero learnt to play the game of Go solely by playing against itself from completely random play~\cite{deepmind2017}. And after 3 days of training, its performance surpassed the performance of the previous version trained by the recorded game of historical pro players (its database holds 30 million moves)~\cite{chen2016}. Another stunning truth unveiled by the AI AlphaGo Zero is that although human invented Go for about 3,000 years and dominated Go for several decades even after the commencement of computer artificial intelligence, AlphaGo Zero has found that we humans seem just found some local optimum strategy of this game in the past 3,000 years. The original version of AlphaGo probably achieved the best performance at its database, which made it become the best Go AI played with a human style. However, AlphaGo Zero proved that the humans played the game in a poor way for more than 3,000 years. That is to say, for some problems the existed human expertise and data (human knowledge) may not be as reliable as we expected~\cite{alphagozero2017}. 

Similar things could occur in the field of software development. What we intuitively think to be right can actually be wrong, or at least, not as good as we supposed them to be. It is hard to achieve the global optimum for humans when the problem is either too large or too complicated, however, such barriers do not exist for artificial intelligence algorithms as long as the algorithms are well designed and enough computing resources have been provided. Therefore, artificial intelligence can really help us detect the bottleneck of our software process and find the global optimum solution instead of local optimum solution.

The breakthrough ability to learn from scratch made this artificial intelligence approach has a potential to be utilised in other fields. If we can generalise the approach used in AlphaGo, it can be suitable for many problems that share similar properties with Go, such as task planning problems in software development or problems where a sequence of actions has to be taken in the right order. Suppose we want to estimate the cost and locate resources of a software project.
Another thing that is very inspiring for us is the iterative development method, which used by DeepMind to develop and evolve their AlphaGo AI program. 

\section{Comparison and Discussions}

IBM Watson integrates several AI techniques such as machine learning, NLP and statistical analysis to analyze and understand the given questions~\cite{gliozzo2013}. The unstructured natural language such as text, image or speech is analyzed and interpreted into structured information which is well-defined and explicit using these AI techniques~\cite{ferrucci2012}. The possible answers generated by Watson are compared based on the rank of confidence values, and then Watson will give a response to the given question. The whole process takes only about 3 seconds. The AI techniques used in Watson are very powerful in solving real-world problems based on the learning ability. The machine learns from the interactions between humans and data, and it can even adapt and become much smarter through the learning process gradually~\cite{deloitte2015}. In contrast, Google AlphaGo uses artificial neural networks to simulate human neural architecture to learn from the best human game players~\cite{granter2017}. AlphaGo also played against itself in order to surpass the best humans, it learned and improved again and again based on reinforcement learning to determine which move contributes to better results~\cite{granter2017}. 

It seems that AI techniques used in AlphaGo are more promising and useful in medical research such as diagnosis and treatment planning. Deep learning systems in AlphaGo can analyze complex information from unstructured data, and the simulated neurons in brain-inspired architecture are strengthened based on the learning process. Moreover, the decision-making system which is very similar to the neurotransmitter dopamine reward system in human brain learned from numerous trials so that it becomes master in some areas~\cite{gibney2015}. However, AlphaGo is powerful might because of its advancement in computing and powerful storage, which are the weakest aspects of human beings compared to computers~\cite{chen2016}. It would be more interesting to see the outcomes if its huge database is removed. To surpass other contestants at Jeopardy, one should have enough knowledge, understand the language, think and respond quickly. This seems much easier than playing chess; however, natural languages are ambiguous, not well-defined and have implicit meanings, whereas chess is a well-defined game which is very complete, explicit and mathematic-based~\cite{ferrucci2012}. 

Comparing IBM Watson and Google AlphaGo is like comparing orange and apple since it is hard to make comparisons between them directly and they are targeting at different application areas. Watson focuses on the commercialization area, whereas AlphaGo focuses on the academic area. There is a big difference between two applications of AI techniques: AlphaGo uses `self-decision making' (make decisions itself) to choose the next move which may contribute to a better outcome, whereas Watson adopts `decision making support' to help human to make decisions~\cite{tak2017}. Another difference is that Watson and AlphaGo have different hardware configurations to support their AI operations: IBM Watson uses a supercomputer, whereas AlphaGo uses multiple distributed smaller computers such as PCs~\cite{tak2017}. Both of them use deep learning and machine learning techniques for training, but Watson uses man-made resources such as papers to learn, AlphaGo, on the other hand, makes full use of massive online resources such as images to learn. IBM company trains the Watson to specific requirements based on customer needs (business to business); however, Google uses more data to create generic techniques (business to customer)~\cite{tak2017}.

SE and AI are two different areas in computer science and both areas have different features, benefits and limitations. SE aims at learning, designing, implementing and improving the software product so that the product can be built within limited time and budget. However, a lot of issues exposed in terms of productivity, reliability and maintenance of software~\cite{jain2011}. Simulating human behaviour or mind using SE is very difficult, and the consciousness of computer is impossible in SE. In addition, most of software process models use fixed phases and sequential methods which leads to inflexible software in nature~\cite{jain2011}. In contrast, AI aims at studying and designing intelligent agents to create intelligent machines. Moreover, it tries to solve problems such as realizing and implementing human mind, emotions, behaviour and intelligence in software products~\cite{jain2011}. Since SE and AI share information with each other and both areas have its own benefits and limitations, which creates more ideas and possibilities in research area, AI methods and techniques have been applied to SE so that the benefits are sum up and limitations are reduced~\cite{jain2011}. Based on two case studies of applying AI techniques in SE, the intersection between both areas is created and developed, and more and more researchers are trying to explore the framework, relation and communication between these two areas. SE uses project management strategy to control and manage software product in each stage to create artificial intelligent software systems; whereas AI adopts knowledge acquisition and data analysis techniques to build intelligent systems. These AI techniques give support to the construction of software products. For example, KBS and CBR are used for requirements engineering during software development process.

Adopting AI techniques such as deep learning, machine learning and business rules in software development may contribute to smarter applications such as Watson and AlphaGo. Integrating AI techniques in software development process can help developers build high quality software faster. On the other hand, the developers need to master these core AI techniques so that they can be used to build software systems that learn on their own. However, computer can understand nothing although it can calculate almost anything, and this one of the limitations to view computers as thinkers. It is very difficult to mimic the human brain since human brain is the most supercomputer and it is different from hardware to software and everything between them~\cite{kachur2017}. Apart from that, human brain is a self-organizing system which is a parallel machine, whereas computers are serial and modular~\cite{chatham2007}. Based on these two case studies, using AI techniques in software systems contributes to more intelligent systems. It would be more promising to see that computers and software systems will become more and more intelligent to help human deal with real-world challenging problems that human beings cannot do.

\section{Conclusion}

AI and SE are two branches of computer science, and both have different strengths and weaknesses. SE focuses on analysing, designing, developing and deploying robust and reliable software systems; whereas AI aims at making systems or machines much smarter by reproducing some aspects of human intelligence. AI techniques such as KBS, NNs and NLP have been applied in software development to deal with challenging problems like over budget, late delivery and software failures in SE.

In this research, two case studies which are IBM Watson and Google AlphaGo that uses AI techniques to solve real world challenging problems have been analysed, evaluated and compared. Based on the analysis of both case studies, using AI such as deep learning and machine learning in software systems contributes to intelligent systems. These intelligent systems continuously learn on their own and make decisions for new problems. The difference between Watson and AlphaGo is that Watson uses `decision making support' to help people make decisions; whereas AlphaGo adopts `self-decision making' to choose operations that contribute to the best outcome. In addition, Watson learns from man-made resources; AlphaGo, on the other hand, learns from massive online resources such as photos. AlphaGo uses neural networks and reinforcement learning to mimic human brain to make decisions intelligently, which might be very useful in medical research for diagnosis and treatment. However, there is still a long way to go if we want to reproduce human brain in machine and view computers as thinkers, as human brain and machines are intrinsically different. It would be more promising to see if AI techniques can be used in software development to deal with challenging problems that manpower cannot solve in the future.

\section{Acknowledgements}

We are grateful to Jiaqi Han, Jiaojie Wei, Fangpeng Li and Minyang Zhang for their fruitful inspiration.

\pagebreak

\bibliographystyle{IEEEtran}
\bibliography{cits5502research}

\end{document}